\documentclass[submission, Phys]{SciPost}
\newcommand{\prb}{Phys. Rev. B }
\newcommand{\pra}{Phys. Rev. A }
\newcommand{\prl}{Phys. Rev. L }
\newcommand{\nat}{nature}
\usepackage{graphicx}
\usepackage{amsmath}
\usepackage{amstext}
\usepackage{amssymb}%
\usepackage{amsfonts}
\usepackage{hyperref}\usepackage{url}
\usepackage{amsbsy}
\usepackage{cleveref}
\usepackage{booktabs}
\usepackage{caption}
\usepackage{siunitx}
\usepackage{mathrsfs}
\usepackage{amsfonts}
\usepackage{amsbsy}
\usepackage{multirow}
\newtheorem{theorem}{Theorem}

\newtheorem{corollary}{Corollary}
\newtheorem{proof}{Proof}[section]
\begin{document}
\begin{center}{
\Large \textbf{
Entanglement, Non-Hermiticity and Duality
}
}
\end{center}

\begin{center}
Li-Mei Chen\textsuperscript{1},
Shuai A. Chen\textsuperscript{2*},
Peng Ye\textsuperscript{1**}
\end{center}

\begin{center}
{\bf 1} School of Physics and State Key Laboratory of Optoelectronic Materials and
Technologies, Sun Yat-sen University, Guangzhou, 510275, China
\\
{\bf 2} Institute for Advanced Study, Tsinghua University, Beijing,
100084, China
\\
* s-chen16@mails.tsinghua.edu.cn \\
** yepeng5@mail.sysu.edu.cn
\end{center}

\begin{center}
\today
\end{center}

\section*{Abstract}
{ Usually duality process keeps energy spectrum invariant. In this paper, we provide a duality, which keeps entanglement spectrum invariant, in order to diagnose  quantum entanglement of   non-Hermitian non-interacting fermionic systems. We limit our attention to non-Hermitian systems with    a  complete set of biorthonormal eigenvectors and an entirely real energy spectrum. 
The original system has a reduced density matrix $\rho_\mathrm{o}$ and the real space is partitioned via a  projecting operator $\mathcal{R}_{\mathrm o}$. After dualization, we obtain a
new reduced density matrix $\rho_{\mathrm{d}}$ and a new real space projector $\mathcal{R}_{\mathrm d}$. Remarkably, entanglement spectrum and entanglement entropy keep invariant. Inspired by the duality, we
defined two types of non-Hermitian models, upon $\mathcal R_{\mathrm{o}}$ is given.
In type-I exemplified by the ``non-reciprocal   model'', there exists
at least one duality such that $\rho_{\mathrm{d}}$ is Hermitian. In other
words, entanglement information of type-I non-Hermitian models with a given   $\mathcal{R}_{\mathrm{o}}$  is entirely controlled by Hermitian models
with  $\mathcal{R}_{\mathrm{d}}$. As a result, we are
allowed to apply known results of Hermitian systems  to efficiently obtain
entanglement properties of type-I models. On the other hand,
the duals of type-II models, exemplified by ``non-Hermitian
Su-Schrieffer-Heeger model'', are always non-Hermitian. For the practical purpose, the
duality provides a potentially \textit{efficient} computation route to   entanglement of
non-Hermitian systems. Via connecting different models, the duality also sheds lights on either trivial or nontrivial role of
non-Hermiticity played in quantum entanglement, paving the way to potentially systematic 
classification and characterization of non-Hermitian systems from the
entanglement perspective. 
}

\vspace{10pt}
\noindent\rule{\textwidth}{1pt}
\tableofcontents\thispagestyle{fancy}
\noindent\rule{\textwidth}{1pt}
\vspace{10pt}

\tableofcontents

\section{Introduction}

For the past decades, Hermitian quantum matters have been intensively
investigated. Classification and characterization of Hermitian quantum
matters are deeply rooted in many-body treatment on quantum entanglement.
Without symmetry, gapped phases are classified into short-ranged entangled
(SRE) and long-range entangled (LRE) phases \cite{wen2004quantum,
2010PhRvBChenLocal}. LRE phases, such as fractional quantum Hall states \cite%
{1983PhRvLLaughlinAnomalous} are physically characterized by the robust
ground state degeneracy on closed manifold and braiding statistics of
topological excitations. Such phases are often called \emph{intrinsic} topological
order \cite{2006KitaevAnyons, 10.1093/nsr/nwv077, 2017RMPWenZoo,
2008AnPhy.323.2709B, 2018PhRvLChanBraiding}. LRE phases cannot be
adiabatically connected to a direct product state via a local unitary
transformation (LU) that attempts to disentangle local degrees of freedom.
In contrast to LRE, there exists \emph{at least} one LU transformation such
that SRE states can be connected to the direct product state without
crossing phase transitions. When symmetry is considered, both LRE and SRE
have finer phase structures. Symmetry Protected Topological phases, e.g.,
the Haldane phase \cite{1983HaldaneContinuum, 1987AffleckRigorous}, are
symmetric SRE states that admit symmetry-protected boundary anomaly  \cite%
{2009PhRvBGuTensor, 2012PhRvBPollmannSymmetry, 2012SciChenSymmetry,
2013PhRvBChenSymmetry, 2017RMPWenZoo}. On the other hand, symmetric LRE
states are called Symmetry Enriched Topological phases \cite%
{2017RMPWenZoo,2019PhRvBBarkeshliSymmetry,2018arXiv180101638N,ye16_set} that
admit fractionalized quantum number carried by topological excitations.
Inspired by quantum information, by partitioning the real-space ${X}$ into
two subregions: ${X_\mathrm o}=\mathcal{A}_\mathrm o \cup \mathcal{B}_\mathrm o$, quantum entanglement
between the two subregions can be quantitatively measured via von Neumann
entanglement entropy (EE): $S_{\mathrm{EE}}=-\mathrm{Tr}\rho \log \rho $
with $\rho=:e^{-h^\mathrm{E}}$ being a reduced density matrix of the
subregion $\mathcal{A}_\mathrm o$ \cite{entanglementreview2009}.
 The full spectrum of
the entanglement ``Hamiltonian'' $h^\mathrm{E}$,  known as  entanglement
spectrum (ES)  \cite{2008HaldaneEntanglement} encodes more fruitful
information about quantum entanglement. In short, it has been well recognized that EE and ES can help
identify and distinguish universal properties of phases \cite%
{2006KitaevAnyons, 2006LevinDetecting, 2006KitaevTopological,
2010PhRvBChenLocal, 2012QiGeneral}.

On the other hand, Hermiticity of a Hamiltonian is one of the key postulates
of isolated quantum systems in order to ensure both probability conservation
and the real-valuedness of eigen-energies. Nevertheless, non-Hermiticity is still
physically relevant and ubiquitous in, e.g., open systems. Non-Hermitian
physics provides a versatile platform for a variety of classical and quantum
systems with concrete lattice models such as non-reciprocal model, non-Hermitian
SSH model \cite{2020arXiv200601837A,1980SSH,HNmodel,2013PhRvALiang,
2017PhRvAKlettRelation, 2018LieuTopological, 2018EPJDKlett,
2018YinGeometrical, 2018YaoEdge,2019MuEmergent}. 

For non-Hermitian systems, while there has been tremendous progress in many
aspects, the entanglement information is far less known, compared to the
progress on many-body entanglement of Hermitian quantum matters such as aforementioned  LRE
and SRE. One may ask whether or not the introduction of non-Hermiticity can
substantially reshape universal behaviors of entanglement properties of
Hermitian systems \cite{entanglementreview2009, 2015JPhA48dFT01B, 2016JPhA49o4005B,2017PhRvL119d0601C}. One may alternatively ask
whether or not there exist non-Hermitian systems whose ES 
can be computed from well-studied Hermitian systems?
   In other words, 
non-Hermiticity  in such systems is irrelevant in the quantum entanglement.
Finally, is it
possible to unify non-Hermitian and Hermitian systems from the entanglement
perspective? 

While general correlated systems are difficult, let us focus on
non-Hermitian non-interacting systems with Hamiltonian\textit{\ }$H_{\mathrm{o}}
$\footnote{If such non-Hermitian systems are realized as meanfield Hamiltonians of correlated systems, the results in this paper are also applicable.}\cite{1960JMP1409W,PhysRevLett805243,2002JMP43205M, 2002JMP432814M,2007RPPh70947B, 2010IJGMM071191M,bender2018pt}.
Then we can construct a reduced density matrix $\rho _{\mathrm{o}}$ by
pairs of right and left eigenstates of $H_{\mathrm{o}}$ \cite%
{2020ChangEntanglement}. More specifically, the entanglement Hamiltonian of $%
H_{\mathrm{o}}$\footnote{Unless otherwise  stated, Hamiltonians of non-Hermitian systems in this paper are
assumed to act on the Hilbert space with a complete set of biorthonormal eigenvectors  
and  possess an entirely real spectrum.}, denoted by $h_{\mathrm{o}}^{\mathrm{E}}$, is analytically
determined by $h_{\mathrm{o}}^{\mathrm{E}}=\log [(\mathcal{R_{\mathrm{o}}P_{%
\mathrm{o}}R_{\mathrm{o}}})^{-1}-\mathbb{I}]$, where two operators $\mathcal{%
R}_{\mathrm{o}}$ and $\mathcal{P}_{\mathrm{o}}$ ($\mathcal{R}_{\mathrm{o}%
}^{2}=\mathcal{R}_{\mathrm{o}}\,,\mathcal{R}_{\mathrm{o}}^{\dagger }=%
\mathcal{R}_{\mathrm{o}}\,,\mathcal{P}_{\mathrm{o}}^{2}=\mathcal{P}_{\mathrm{%
o}}$, $\mathcal{P}_{\mathrm{o}}^{\dagger }\neq \mathcal{P}_{%
\mathrm{o}}$) impose quantum-state projections onto the subregion $\mathcal{A%
}_\mathrm o$ of $X_\mathrm o$ and occupied eigenstates of $H_{\mathrm{o}}$, respectively. It
should be noted that the Fock-space projector $\mathcal{P}_{\mathrm{o}}$ 
is no longer Hermitian, but the real-space projector $\mathcal{R}_{\mathrm{o}%
}$, by definition, must always be Hermitian.

\begin{figure}[ht]
\centering 
\includegraphics[scale=0.37]{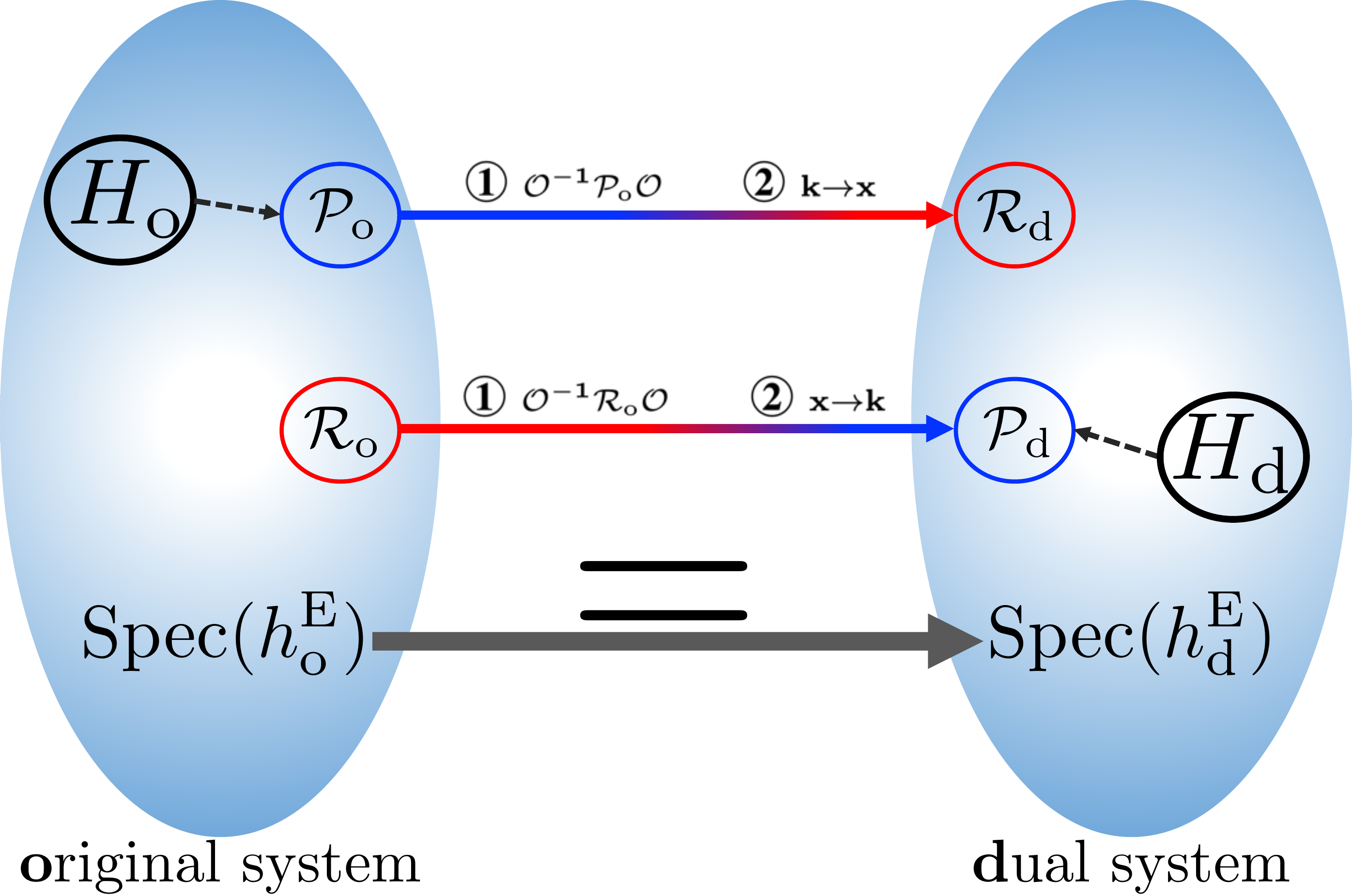} 
\caption{Schematic illustration of the duality.  
The duality between the original system $H_{\mathrm o}$ and the dual system $H_{\mathrm d}$ is split into two steps. 
  In the  first step {\small \textcircled{\small{\textbf 1}}}, we introduce a similarity transformation on both Fock-space projector $\mathcal P_\mathrm o$ and real-space projector $\mathcal R_\mathrm o$.  In the second step {\small \textcircled{\small{\textbf 2}}}, we interchange real space and momentum space.  
  Entanglement spectrum and entanglement entropy, which can be computed from diagonalizing reduced density matrices as in Eq.~\eqref{entdef},
 keep invariant. Details can be found in the main text.  
 }
\label{dual2step}
\end{figure}

In this paper, we build a rigorous duality between a non-Hermitian
non-interacting Hamiltonian $H_{\mathrm{o}}$ and its dual Hamiltonian, denoted
by $H_{\mathrm{d}}$. Remarkably, $H_{\mathrm{o}}$ and $H_{\mathrm{d}}$ share
the same ES, i.e., $\mathrm{Spec}(h_{\mathrm{o}}^{\mathrm{E}})=\mathrm{Spec}%
(h_{\mathrm{d}}^{\mathrm{E}})$.\footnote{Here\ the symbol $\mathrm{Spec}(\mathcal{O})$
denotes the spectrum of the operator $\mathcal{O}$.} Meanwhile, the dual
Hamiltonian $H_{\mathrm{d}}$ and its   reduced density matrix $\rho _{\mathrm{d}}$
may be either non-Hermitian or Hermitian. By means of this duality, we
establish exotic connections between different models, regardless of
Hermiticity. Before moving to detailed technical discussions to appear in the
main text, let us concisely illustrate the duality here via Fig.~\ref{dual2step}. 
In Fig.~\ref{dual2step}, the duality consists of two key steps.  Step-{\small \textcircled{\small{\textbf 1}}}, a similarity
transformation $\mathcal{O}$ is applied to \textit{not only} $\mathcal{P}_{%
\mathrm{o}}$ \textit{but also} $\mathcal{R}_{\mathrm{o}}$.  
The feature of `\textit{simultaneously acting on both projectors}' 
is very crucial and will be elaborated in the main text.  Step-{\small \textcircled{\small{\textbf 2}}}, the real
space and Fock-space are exchanged.
 As a result, two new projectors $%
\mathcal{R}_{\mathrm{d}}$ and $\mathcal{P}_{\mathrm{d}}$ of the dual system
are naturally defined. Both projectors enter $h_{\mathrm{d}}^{\mathrm{E}}$
in the standard way: $h_{\mathrm{d}}^{\mathrm{E}}=\log [(\mathcal{R_{\mathrm{%
d}}P_{\mathrm{d}}R_{\mathrm{d}}})^{-1}-\mathbb{I}]$. Since $\mathcal{R}_{%
\mathrm{d}}$ is interpreted as a real-space projection in the dual system, its
Hermiticity must be guaranteed: $\mathcal{R}_{\mathrm{d}}^{{}}=\mathcal{R}_{%
\mathrm{d}}^{\dagger }$. Therefore, we require that the equality $\Theta \mathcal{P}_{%
\mathrm{o}}^{\dagger }=\mathcal{P}_{\mathrm{o}}\Theta $ always holds, where $%
\Theta :=\mathcal{O}\mathcal{O}^{\dag }$. Theorem~\ref{Thm1} of the main
text will be introduced to guarantee $\mathrm{Spec}(\mathcal{R}_{\mathrm{o}}%
\mathcal{P}_{\mathrm{o}}\mathcal{R}_{\mathrm{o}})=\mathrm{Spec}(\mathcal{R}_{%
\mathrm{d}}\mathcal{P}_{\mathrm{d}}\mathcal{R}_{\mathrm{d}})$. Therefore,
the duality in Fig.~\ref{dual2step} keeps ES unaffected, i.e., $\mathrm{Spec%
}(h_{\mathrm{o}}^{\mathrm{E}})=\mathrm{Spec}(h_{\mathrm{d}}^{\mathrm{E}})$, which completes the duality process.

The duality has profound consequences. Physically, the duality inspires us
to divide non-Hermitian Hamiltonians into two types, namely, type-I and
type-II, when a real-space partition $\mathcal{R}_{\mathrm{o}}$ is given. In
each case of type-I, there exists at least one duality process such that $%
\mathcal{P}_{\mathrm{d}}^{\dagger }=\mathcal{P}_{\mathrm{d}}^{{}}$ and thus $%
\rho _{\mathrm{d}}^{\dagger }=\rho _{\mathrm{d}}^{{}}$. Therefore, in
type-I, ES can be fully determined by its Hermitian dual, i.e., $H_{\mathrm{d%
}}$. On the contrary, in type-II, it is impossible to compute ES from known
results of Hermitian systems through the present duality, since all $H_{\mathrm{%
d}}$'s are non-Hermitian. Mathematically, whether $\rho _{\mathrm{d}}$ is
Hermitian or not essentially relies on the commutator $[\Theta ,\mathcal{R}_{%
\mathrm{o}}]$. If there exists an $\mathcal{O}$ such that $[\Theta ,\mathcal{%
R}_{\mathrm{o}}]=0$, then $\rho _{\mathrm{d}}$ is Hermitian and $H_{\mathrm{o%
}}$ is of type-I. On the one hand, from the definition, non-Hermiticity of
type-I doesn't play any essential role in EE and ES. But on the other hand, for the practical purpose, 
we are allowed to \textit{efficiently} compute ES and EE of type-I systems
by means of known results of Hermitian systems. To demonstrate it, the
non-reciprocal lattice model is identified as type-I. As a byproduct, we
prove that ES and EE of this model are independent of the parameter $\alpha $
[Eq.~(\ref{nonreciprocal})] that measures the degree of ``non-reciprocality". In contrast to type-I, the dual system 
$H_{\mathrm{d}}$ is always non-Hermitian for type-II systems. It indicates
that entanglement information of type-II system cannot be understood through
any known results of Hermitian systems in the present duality process. The
non-Hermitian Su-Schrieffer-Heeger (SSH) model is one of simplest examples
of type-II.	In Table~\ref{Tab-1}, we list the criteria and entanglement properties of 
	the two types of systems.

The remainder of this paper is organized as follows. In Sec.~\ref{section_duality_preli}, some useful facts on non-Hermitian quantum physics are reviewed. We explain the duality process by presenting
one theorem and two corollaries in Sec.~\ref{section_duality_process}. Two
typical examples (nonreciprocal model and non-Hermitian SSH model) are
computed in details in Sec.~\ref{section_two_typical_examples}. In Sec.~\ref%
{section_conclusions}, this work is concluded with several remarks and
future directions. Appendices include further supplemental information on
the duality.

\section{Preliminaries}\label{section_duality_preli}

For a non-Hermitian system of free fermions, its second-quantized Hamiltonian can be written as%
\begin{equation}
H_{\mathrm{o}}=\sum_{\alpha \beta }c_{\alpha }^{\dag }H_{\alpha \beta
}c_{\beta }^{{}}\,,
\end{equation}%
where $H_{\mathrm{o}}\not=H_{\mathrm{o}}^{\dag }$ and fermionic operators $%
c_{\alpha }^{\dag }$ and $c_{\alpha }$ satisfy the standard anticommunication relations $%
\{ c_{\alpha }^{\dag },c_{\beta }\} =\delta _{\alpha \beta }$.
Suppose $H_{\mathrm{o}}$ admits a complete set of biorthonormal eigenvectors 
$\{| r,\alpha \rangle ,| l,\alpha \rangle \}$ that satisfy 
\begin{equation}
\langle  l,\alpha | r,\beta \rangle =\delta _{\alpha \beta }~,\quad
\sum_{\alpha }| l,\alpha \rangle \langle  r,\alpha |=\mathbb{I}
\end{equation}%
with $| r,\alpha \rangle $ and $| l,\alpha \rangle $ being the right and left
eigenvectors,  
\begin{equation}
H_{\mathrm{o}}| r,\alpha \rangle =\epsilon _{\alpha }| r,\alpha \rangle
~,\quad H_{\mathrm{o}}^{\dagger }| l,\alpha \rangle =\epsilon _{\alpha
}^{\ast }| l,\alpha \rangle ~.
\end{equation}%
Here $\alpha $ is the spectral label, $\delta _{\alpha \beta }$ denotes the
Kronecker delta function and $\mathbb{I}$ is the identity matrix. Therefore,
we have a spectral decomposition 
\begin{equation}
H_{\mathrm{o}}=\sum_{\alpha }\epsilon _{\alpha }| r,\alpha \rangle \langle
 l,\alpha |~.
\end{equation}%
By introducing bifermionic operators 
\begin{equation}
\psi _{ r\alpha }^{\dag }|0\rangle \equiv | r,\alpha \rangle ~,\quad \psi
_{ l\alpha }^{\dag }|0\rangle \equiv | l,\alpha \rangle ~,
\end{equation}%
with the anti-commutation relations $\{\psi _{ l\alpha }^{{}},\psi _{ r\beta
}^{\dag }\}=\delta _{\alpha \beta }$, we can straightforwardly construct a many-body state, 
\begin{equation}
|G_{ r}\rangle =\prod_{\alpha \in \text{occ}}\psi _{ r\alpha }^{\dag
}|0\rangle ~,\quad |G_{ l}\rangle =\prod_{\alpha \in \text{occ}}\psi
_{ l\alpha }^{\dag }|0\rangle ~,  \label{mbstate}
\end{equation}%
where occ denotes a set of the occupied states. \emph{Hereafter, in the
rest of the paper, unless otherwise  stated, a non-Hermitian Hamiltonian is assumed to 
act on the Hilbert space with a complete set of  biorthonormal eigenvectors and 
possess an
entire real energy spectrum.}
Mathematically, it is equivalent to the condition that 
\cite{2002JMP43205M, 2002JMP432814M}
there is an invertible operator $\mathcal{O}$ such that 
\begin{equation}
H_{\mathrm{o}}\Theta =\Theta H_{\mathrm{o}}^{\dagger }~,  \label{ho=ohd}
\end{equation}%
with 
\begin{equation}
\Theta :=\mathcal{O}\mathcal{O}^{\dagger}~.  \label{Theta}
\end{equation}

From the right and left states, a density matrix can be constructed \cite%
{2020ChangEntanglement} as
\begin{equation}
\rho =|G_{r}\rangle \langle G_{l}|
\end{equation}%
such that $\rho ^{2}=\rho $ and $\rho ^{\dag }\not=\rho $. With this
generalized notation, provided a partition on the real-space into subregions 
$X_\mathrm o=\mathcal{A}_\mathrm o\cup \mathcal{B}_\mathrm o$, we realize measurement of entanglement 
$S_{\mathrm{EE}}=-\mathrm{Tr}\rho _{\mathrm o}\ln \rho _{\mathrm o}$,
where the
reduced density matrix $\rho_{\mathrm o}$ is defined by taking partial
trace of subsystem $\mathcal{B}_\mathrm o$, 
\begin{equation}
\rho _{\mathrm o}=\mathrm{Tr}_{\mathcal{B}_\mathrm o}\rho =:e^{-h_\mathrm{o}^{\mathrm{E}}},~h_\mathrm{o}^{%
\mathrm{E}}=\sum_{i,j\in \mathcal{A}_\mathrm o}c_{i}^{\dag }(h_{\mathrm{E}%
})_{ij}c_{j}~.  \label{entdef}
\end{equation}%
The entanglement Hamiltonian $h_\mathrm{o}^{\mathrm{E}}$ is introduced \cite%
{2008HaldaneEntanglement} in Eq.~(\ref{entdef}) whose spectrum $\mathrm{Spec}%
(h_\mathrm{o}^{\mathrm E})$ encodes more fruitful information on quantum entanglement. For a
non-interacting system, entanglement Hamiltonian $h_\mathrm{o}^{\mathrm{E}}$ is uniquely
determined by a two-point correlation matrix $C_{\mathrm{o}}$ with elements $%
(C_{\mathrm{o}})_{ij}=\langle G_{l}|c_{i}^{\dag }c_{j}|G_{r}\rangle $, $%
i,j\in \mathcal{A}_\mathrm o$ via a relation 
\begin{equation}
h^{\mathrm{E}}_\mathrm o=\log \left( C_{\mathrm{o}}^{-1}-\mathbb{I}\right) 
\label{hec}
\end{equation}%
with $\mathbb{I}$ being an identity matrix \cite{2003PeschelCalculation,
2009PeschelReduced}. Furthermore, we can reformulate $C_{\mathrm{o}}$ as 
\begin{equation}
C_{\mathrm{o}}=\mathcal{R}_{\mathrm{o}}\mathcal{P}_{\mathrm{o}}\mathcal{R}_{%
\mathrm{o}}~,  \label{cm}
\end{equation}%
in terms of the real-space projector 
\begin{equation}
\mathcal{R}_{\mathrm{o}}=\sum_{i\in \mathcal{A}_\mathrm o}|i\rangle \langle i|
\end{equation}%
onto region $\mathcal{A}_\mathrm o$ and Fock-space projector 
\begin{equation}
\mathcal{P}_{\mathrm{o}}=\sum_{\alpha \in \mathrm{occ}}|r,\alpha \rangle
\langle l,\alpha |
\end{equation}%
onto occupied states \cite{2014LeePosition, 2020ChangEntanglement}. One
significant feature is that $\mathcal{P}_{\mathrm{o}}$ is no longer
Hermitian, namely, $\mathcal{P}_{\mathrm{o}}\not=\mathcal{P}_{\mathrm{o}%
}^{\dag }$, while the real-space projector $\mathcal{R}_{\mathrm{o}}$ is, by
definition, Hermitian $\mathcal{R}_{\mathrm{o}}=\mathcal{R}_{\mathrm{o}%
}^{\dag }$.
In our duality [see Fig.~\ref{dual2step}], the basic notations for the dual system 
can be obtained by replacing the subscript index $\mathrm{o}$ with $\mathrm{d}$. For example, 
the reduced density matrix $\rho_\mathrm{d}$ for the dual system defines its entanglement Hamiltonian $h_\mathrm{d}^\mathrm{E}$ via $\rho_\mathrm{d}=:e^{-h_\mathrm{d}^\mathrm{E}}$
and other formula work in the same way.

\section{Duality}

\label{section_duality_process} 
When entanglement meets non-Hermiticity,  how are the universal 
behaviours of entanglement  reshaped? Alternatively, is it possible that 
we can comprehend entanglement of non-Hermitian systems based on the knowledge of 
Hermitian systems?
For this purpose, we propose a duality, which is depicted in Fig.~\ref{dual2step}. This duality process keeps ES and EE unaffected and leads to two different types of non-Hermitian models.


\subsection{Duality process}
 
As schematically illustrated in Fig.~\ref{dual2step},
our duality is conducted by two steps.
In the first step, under a similarity transformation,   $\mathcal{
P}_{\mathrm{o}}$ is mapped to a Hermitian operator, and, \textit{simultaneously} $\mathcal R_\mathrm o$ is 
mapped to an operator that may or may not be Hermitian.
We exchange the roles of momenta and positions in the second step and obtain projectors $\mathcal P_\mathrm d $ and $\mathcal R_\mathrm d$ in a dual system.   
Since a projector $\mathcal R_\mathrm d$ must be Hermitian $\mathcal{R}_\mathrm{d}^\dagger=\mathcal{R}_\mathrm{d}$ in order to depict a real-space partition, we have to impose an requirement on  the similarity transformation  $\mathcal O$ in the first step, 
\begin{equation}
\Theta \mathcal{P}_\mathrm{o}^\dagger = \mathcal P_\mathrm o \Theta
\label{TppT}
\end{equation} 
with $ \Theta =\mathcal{O}\mathcal{O}^\dagger$ defined in Eq.~\eqref{Theta}. 
Besides, invariance of ES and EE requires a condition 
\begin{equation}
\mathrm{Spec}(\mathcal{R}_\mathrm{o} \mathcal{P}_\mathrm{o}\mathcal{R}_\mathrm{o})=\mathrm{Spec}(\mathcal{R}_\mathrm{d} \mathcal{P}_\mathrm{d}\mathcal{R}_\mathrm{d})~,
\label{spec=spec}
\end{equation}
as indicated by  Eq.~\eqref{hec}.
The two conditions in Eqs.~\eqref{TppT} and \eqref{spec=spec} can be satisfied as indicated by Theorem~\ref{Thm1}.

\begin{theorem}
\label{Thm1} Given a Hamiltonian $H_{\mathrm{o}}$ acting on the Hilbert
space with a complete set of biorthonormal eigenvectors with an entirely real
spectrum, there exists an invertible similarity transformation 
$\mathfrak{A}=\mathcal{O}^{-1}\mathcal{P}_{\mathrm{o}}\mathcal{O}$ and $%
\mathfrak{B}=\mathcal{O}^{-1}\mathcal{R}_{\mathrm{o}}\mathcal{O}$
such that 
\begin{equation}
\mathrm{Spec}\left( \mathcal{R}_{\mathrm{o}}\mathcal{P}_{\mathrm{o}}\mathcal{%
R}_{\mathrm{o}}\right) =\mathrm{Spec}( {\mathfrak{ABA}}) 
\label{sp=sp}
\end{equation}%
with $\mathfrak{A}=\mathfrak{A}^{\dag }$. Here the
symbol $\mathrm{Spec}(\mathcal{O})$ in Eq.~\eqref{sp=sp} denotes spectrum of $%
\mathcal{O}$.
\end{theorem}

Theorem~\ref{Thm1} states that there is always an invertible similarity transformation 
such that a rearrangement on projectors keep the spectrum unaffected, which is a generalized version 
of the Hermitian counterpart \cite{2012arXiv1205.6266H, 2014LeePosition}.
We can exchange roles of momenta and positions,  in the
second step. We re-interpret $\mathfrak{B}$ as a new Fock-space
projector, \textit{re-denoted as}   $\mathcal{P}_{\mathrm{d}}$, to describe occupied
states and $\mathfrak{A}$, as a new real-space projector, \textit{re-denoted
as} $\mathcal{R}_{\mathrm{d}}$ to depict the real partition. Thus, we obtain
a correlation matrix $C_{\mathrm{d}}=\mathcal{R}_{\mathrm{d}}\mathcal{P}_{%
\mathrm{d}}\mathcal{R}_{\mathrm{d}}$ from which we can design the dual
Hamiltonian $H_{\mathrm{d}}$ with an entirely real spectrum. Theorem~\ref{Thm1} along with the formula in
Eq.~\eqref{hec} then indicates that $H_{\mathrm{d}}$ shares the same EE and
ES with $H_{\mathrm{o}}$. At this stage, we finish building our duality
between two systems, $H_{\mathrm{o}}$   with a real-space partition $\mathcal R_\mathrm o$ 
and $H_{\mathrm{d}}$   with a real-space partition $\mathcal R_\mathrm d$.
  The procedures
are depicted in Fig.~\ref{dual2step}. 
Such a duality allows us to diagnose
entanglement properties of a non-Hermitian system $H_{\mathrm{o}}$ from the
computation in  its dual one.

Below we give a proof of Theorem~\ref{Thm1}. 

\begin{proof}

From the property of $H_\mathrm o$,
there is  an invertible operator $\mathcal O$ with $\Theta:=\mathcal{O}\mathcal{O}^\dagger$
that satisfies the condition in Eq.~\eqref{ho=ohd}.
Then, we have 
$
 \Theta \mathcal{P}_\mathrm o^\dagger=\mathcal{P}_\mathrm o\Theta
$,
or equivalently,
\begin{equation}
\mathcal{O}^{-1}\mathcal{P}_\mathrm o \mathcal O=\mathcal{O}^{\dagger}\mathcal{P}^\dagger_\mathrm o \mathcal O^{\dagger-1}~.
\end{equation}
We define $\mathfrak{A}=
\mathcal{O}^{-1}\mathcal{P}_{\mathrm{o}}\mathcal{O}%
 $ and $\mathfrak{A}$ is Hermitian $\mathfrak{A}=\mathfrak{A}^\dagger $.
Since an invertible similarity transformation does not alter spectrum, we have
\begin{equation}
\mathrm{Spec}\left( \mathcal{R}_{\mathrm{o}}\mathcal{P}_{\mathrm{o}}\mathcal{R}_{\mathrm{o}}\right)
=\mathrm{Spec}(\mathcal{O}\mathfrak{BAB}\mathcal{O}^{-1})=
\mathrm{Spec}(\mathfrak{BAB})\,,
\end{equation}
where $\mathfrak{B}=\mathcal{O}%
^{-1}\mathcal{R}_{\mathrm{o}}\mathcal{O}$ may be either  Hermitian or not.

The next is to prove $\mathrm{Spec}(\mathfrak{BAB})=\mathrm{Spec}(\mathfrak{ABA})$. Suppose an eigenstate $|\xi \rangle $ of $\mathfrak{BAB}$ with 
\begin{equation}
\mathfrak{BAB}|\xi \rangle
=c|\xi \rangle~,
\end{equation}
 and then $\mathfrak{B}|\xi \rangle =|\xi
\rangle $ by observing
\begin{equation}
 c\mathfrak{B}|\xi \rangle =\mathfrak{B}\mathfrak{BAB}|\xi \rangle =c|\xi \rangle ~.
\end{equation}
Thus, $\mathfrak{A}%
|\xi \rangle $ is an eigenstate of $\mathfrak{ABA}$:
\begin{equation}
\mathfrak{ABA}(\mathfrak{A}|\xi \rangle )=c%
(\mathfrak{A}|\xi \rangle)~.
 \end{equation}
Therefore, given an eigenstate $|\xi
\rangle $ of $\mathfrak{BAB}$ with eigenvalue $c$, $\mathfrak{A}|\xi \rangle $ is
an eigenstate of $\mathfrak{ABA}$ with eigenvalue $c$. The converse is also true. Finally, we have
\begin{equation}
\mathrm{Spec}\left( \mathcal{R}_{\mathrm{o}}\mathcal{P}_{\mathrm{o}}\mathcal{R}_{\mathrm{o}}\right) =%
\mathrm{Spec}(\mathfrak{BAB})=\mathrm{Spec}(\mathfrak{ABA})\,.
\end{equation}
\end{proof}

\subsection{Two types of non-Hermitian systems}
\label{section_two_types_Definition}

  The duality shown in Fig.~\ref{dual2step} maps a non-Hermitian system $H_\mathrm o$ into a new one $H_\mathrm d$
while they share the same ES and EE. 
Therefore, we can diagnose entanglement in $H_\mathrm o$ by means of $H_\mathrm d$.
If the dual system $H_\mathrm d$ turns out to be Hermitian, we can assert that non-Hermiticity 
indeed does not play any essential role in entanglement of such a system $H_\mathrm o$. 
The condition for $H_\mathrm d$ being Hermitian, 
i.e., 
	\begin{equation}
	  \mathcal{O}^{-1}\mathcal{R}_\mathrm{o}\mathcal{O}=\mathcal{O}^{\dagger}\mathcal{R}_\mathrm{o}\mathcal{O}^{-1\dagger}~,
	\end{equation}
is that a similarity transformation $\mathcal O$ 
exists such that $\Theta$ commutes with $\mathcal{R}_{\mathrm{o}}$,
\begin{equation}
[\Theta ,\mathcal{R}_{\mathrm{o}}]=0~.
\label{thetar}
\end{equation}
Consequently, given $H_\mathrm o$ if at least such a similarity transformation exists to meet Eq.~\eqref{thetar},
we regard such a system as type-I.  Otherwise, the
system is categorized into type-II. 
 Obviously, a Hermitian Hamiltonian belongs to type-I, for which we
can simply take $\mathcal{O}$ to be an identity matrix.  
The ES and EE of type-I obey the same
tendency as a Hermitian system. Thus we can understand it within the context
of Hermitian systems. For example, we expect that the entanglement formula 
 inspired by Wisdom conjecture \cite{2006PhRvLGioevEntanglement, 2006PhRvABarthelEntanglement,
2006PhRvBLiScaling, 2015PhRvBLaiEntanglement}
still work and EE and ES are directly obtained 
from known results on Hermitian
systems without complicated calculations. On the other hand, non-Hermiticity
is supposed to play an intrinsic role in entanglement of type-II.

The operator $\Theta$ varies for different choices on $\mathcal O$.
Explicitly, given $\mathcal O_1$ that satisfies Eq.~(\ref{ho=ohd}),
the operator $\mathcal{O}_{2}=\mathcal{O}_{1}U_{1}SU_{2}$ also meets Eq.~(\ref{ho=ohd}), but $\Theta$ is changed. Here $U_1$ denotes a unitary transformation
that diagonalizes $\mathcal{O}_1 H_\mathrm{o} \mathcal{O}_1^{-1}$, $U_2$ is an arbitrary unitary matrix and $S$ is an invertible matrix 
that commutes with spectral matrix of $H_\mathrm{o}$. In Appdendix~\ref{App:strans}, 
we give an example to illustrate this point.
In practice, to determine the type of a system, one can start with $\mathcal{O}$
that diagonalizes $\mathcal{O}^{-1}H_{\mathrm{o}}\mathcal{O}=\Lambda $. If $%
[ \mathcal{OO}^{\dag },\mathcal{R}_{\mathrm{o}}] =0$, then it belongs to
type-I. Otherwise, we have to check whether some $S$ exists to solve the
equation $\left[ \mathcal{O}SS^{\dag }\mathcal{O}^{\dag },\mathcal{R}_{%
\mathrm{o}}\right]=0$. In Appendix~\ref{App:strans}, we make more explanation on the procedure to determine the type of a given non-Hermitian model. 

Based on the theorem, two corollaries naturally follow.

\begin{corollary}
\label{crlry1} The ES for type-I non-Hermitian system is
real. A non-Hermitian system with complex ES belongs to
type-II.
\end{corollary}

The Corollary~\ref{crlry1} is a direct consequence of the duality process.
The real-valued ES of a type-I arises from the identical spectrum to a Hermitian
system. The converse-negative proposition of the first part produces the
second argument. We point out that we do not exclude a type-II system
possessing a real ES.

\begin{corollary}
\label{crlry2}  In the case of a Hermitian system, one recovers the
``position-momentum duality''.
\end{corollary}

The Corollary~\ref{crlry2} is obvious since one can simply choose $\mathcal{O%
}$ to be an identity matrix, which exactly recovers the result established
in Ref.~\cite{2014LeePosition}. 
  In Table~\ref{Tab-1}, we summarize the properties of two 
types of non-Hermitian systems as well as two typical examples that will be presented 
in Sec.~\ref{sec:example1} and Sec.~\eqref{sec:example2}.
\begin{table}
\caption{Criteria and entanglement spectrum (ES) for categorizations on non-Hermitian free systems as well as two examples.
We obtain two types of non-Hermitian free systems according to whether at least one $\Theta$ in Eq.~\eqref{Theta} exists to commute 
with a real space partition $\mathcal R_\mathrm o$. 
}
\label{Tab-1}
\centering
\begin{tabular}{llll}
\toprule
           &   Criteria  &      ES     & Example    \\ 
\midrule      
Type-I     & At least one $\Theta$ commutes with $\mathcal{R}_\mathrm{o}$ & Real  & Non-reciprocal model [Sec.~\ref{sec:example1}]  \\ 
Type-II    & No $\Theta$ commutes with $\mathcal{R}_\mathrm{o}$  &  Real or complex  & non-Hermitian SSH model [Sec.~\ref{sec:example2}]  \\ 
\bottomrule
\end{tabular}
\end{table}

\section{Examples}\label{section_two_typical_examples}
In this section,  we present two examples to exemplify the
two types of non-Hermitian systems.

\subsection{An example for Type-I: Nonreciprocal model}
\label{sec:example1}

\begin{figure}
\centering
\includegraphics[scale=0.25]{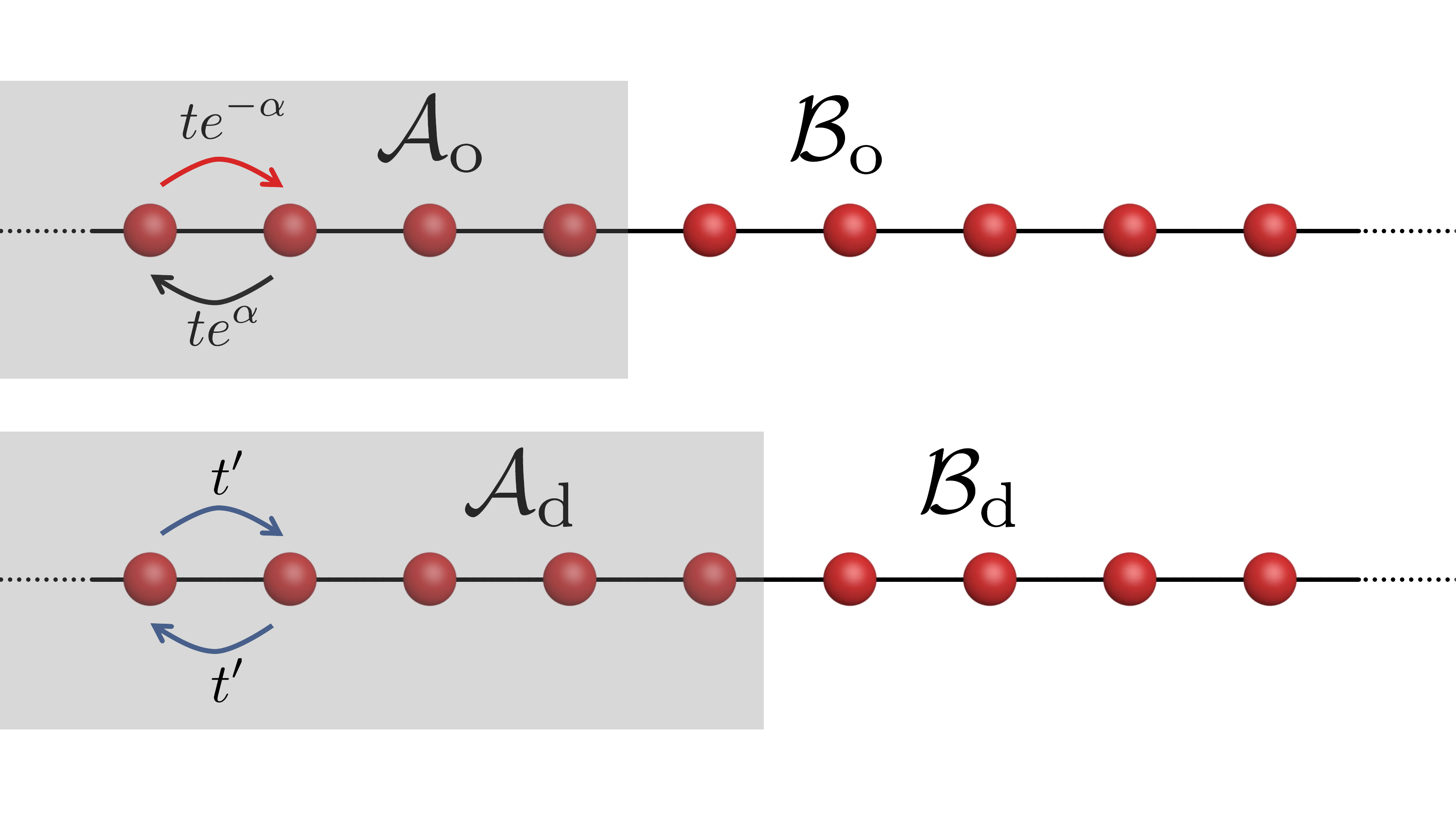}
\caption{Illustration of non-reciprocal model and its dual system for a general filling and partition.
After duality, the partition is changed from $X_\mathrm o=\mathcal A_\mathrm o \cup \mathcal B_\mathrm o$ to 
$X_\mathrm d=\mathcal A_\mathrm d \cup \mathcal B_\mathrm d$. Non-Hermiticity in $H_\mathrm o$ arises from non-reciprocal hopping
$te^{\pm\alpha}$ while the dual system  is Hermitian with hopping integral $t'$. 
}
\label{Fig:nonrecp}
\end{figure}

As a warm-up, we consider one of the simplest non-Hermitian models 
on a chain of $L$ sites 
\begin{equation}
H_{\mathrm{o}}=-t\sum_{x=1}^{L}\left (e^{\alpha }c_{x}^{\dag }c_{x+1}^{}+e^{-\alpha
}c_{x+1}^{\dag }c_{x}^{}\right)\,,  \label{nonreciprocal}
\end{equation}%
where $c_{x}^{\dag }$ and $c_{x}$ are the fermion creation and annihilation
operators at site $x$, respectively. The nonreciprocal left/right hopping $%
te^{\pm \alpha }$ can arise from asymmetric gain/loss, which is shown in Fig.~\ref{Fig:nonrecp}. 
Under an open boundary condition (OBC), it is exactly solvable and one can
write down the right and left eigenvectors $|r,k\rangle $ and $|l,k\rangle $
as 
\begin{align}
| r,k\rangle = & \sqrt{\frac{2}{L+1}}\sum_{x=1}^Le^{-\alpha x}\sin \frac{\pi
kx}{L+1}|x\rangle~, \\
| l,k\rangle = & \sqrt{\frac{2}{L+1}}\sum_{x=1}^Le^{\alpha x}\sin \frac{\pi kx%
}{L+1}|x\rangle~,
\end{align}{}
with a real gapless spectrum $\epsilon_\mathrm o ( k ) =-2t\cos \frac{\pi k}{%
L+1}$ parametrized by momentum indices $k=1,\cdots ,L$. 

Now we conduct the duality in Fig.~\ref{dual2step}.
In \textbf{Step}-\textcircled{\small\textbf{1}}, we can choose $\mathcal O$ to be  
\begin{equation}
\mathcal{O}=\mathrm{diag}(e^{-\alpha},\cdots,e^{-L\alpha})~,
\label{examp1O}
\end{equation}
which describes the transformation 
\begin{equation}
c_{x}^{\dag }\rightarrow c_{x}^{\dag }e^{-x\alpha }~,\quad c_{x}\rightarrow
c_{x}e^{x\alpha }~.
\end{equation}
Partition the system into two subregions $\mathcal{A}_\mathrm o$ and $\mathcal{B}_\mathrm o$ with
the Fock-space and real-space projectors being
\begin{equation}
\mathcal{P}_{\mathrm{o}}=\sum_{k\in \text{occ}}|r,k\rangle \langle l,k|\,,%
\mathcal{R}_{\mathrm{o}}=\sum_{x\in \mathcal{A}_\mathrm o}|x\rangle \langle x|~.
\label{PRex1}
\end{equation}%
After Step-\textcircled{\small\textbf{1}} with a similarity transformation in Eq.~\eqref{examp1O} acting on both $\mathcal P_\mathrm o$  and 
$\mathcal R_\mathrm{o}$ in Eq.~\eqref{PRex1}, 
\begin{equation}
\mathcal O^{-1} \mathcal P_\mathrm o \mathcal O = \sum_{k\in \mathrm{occ}} |k\rangle \langle k|\, , 
\mathcal O^{-1} \mathcal R_\mathrm o \mathcal O = \sum_{x\in \mathcal A_0}|x\rangle \langle x|\, ,
\end{equation}
in \textbf{Step}-\textcircled{\small \textbf{2}}, 
we interchange the roles of momenta and positions, which defines
dual projectors 
\begin{equation}
\mathcal{R}_{\mathrm{d}}=\sum_{x\in \mathrm{occ}}|x\rangle \langle x|\, , 
\mathcal{P}_{\mathrm{d}}=\sum_{k\in \mathcal{A}_\mathrm o}|k\rangle \langle k|\,,%
\end{equation}%
where the two basis's  $|k\rangle$ and $ |x\rangle$ 
in $\mathcal R_\mathrm d$ and $\mathcal P_\mathrm d$ 
satisfy the relation,
$
|k\rangle=\sqrt{\frac{2}{L+1}}\sum_{x=1}^{L} \sin\frac{\pi kx}{L+1}|x\rangle
$.
Here the
Fock-space projector $\mathcal{P}_{\mathrm{d}}$ means that states with
momentum in region $\mathcal{A}_\mathrm o$ are occupied and $\mathcal{R}_{\mathrm{d}}$
defines a real-space partition. 
For example, at half-filling of $H_\mathrm o$, we can conduct a partition with $\mathcal A_\mathrm o$ containing half the chain.
Then we can take the spectral dispersion to be $\epsilon_\mathrm d=-2t^{\prime}\cos\frac{\pi kx}{L+1}$,
and the dual Hamiltonian $H_\mathrm{d}$ reads 
\begin{equation}{}
H_\mathrm d= -t^{\prime}\sum_{x=1}^{L} \left( c_x^\dagger c_{x+1}^{}+c_{x+1}^\dagger c^{}_{x}\right)\, ,
\label{dual1}
\end{equation}
 with a partition $\mathcal A_\mathrm d$ being half the chain \footnote{Here the strength of hopping integral $t^{\prime}$ does not influence entanglement.}. For a general partition, we can introduce a chemical potential such that 
 $\epsilon_\mathrm d ( k ) <0$ when $k \in \mathcal A_\mathrm o$.  In Fig.~\ref{Fig:nonrecp}, we depict lattices for $H_\mathrm o$ and its dual $H_\mathrm d$ with partition $X_\mathrm o=\mathcal A_\mathrm o\cup\mathcal B_\mathrm o$ and $X_\mathrm d=\mathcal A_\mathrm d\cup\mathcal B_\mathrm d$
 respectively.

\begin{figure}[t]
\centering
\includegraphics[scale=0.6]{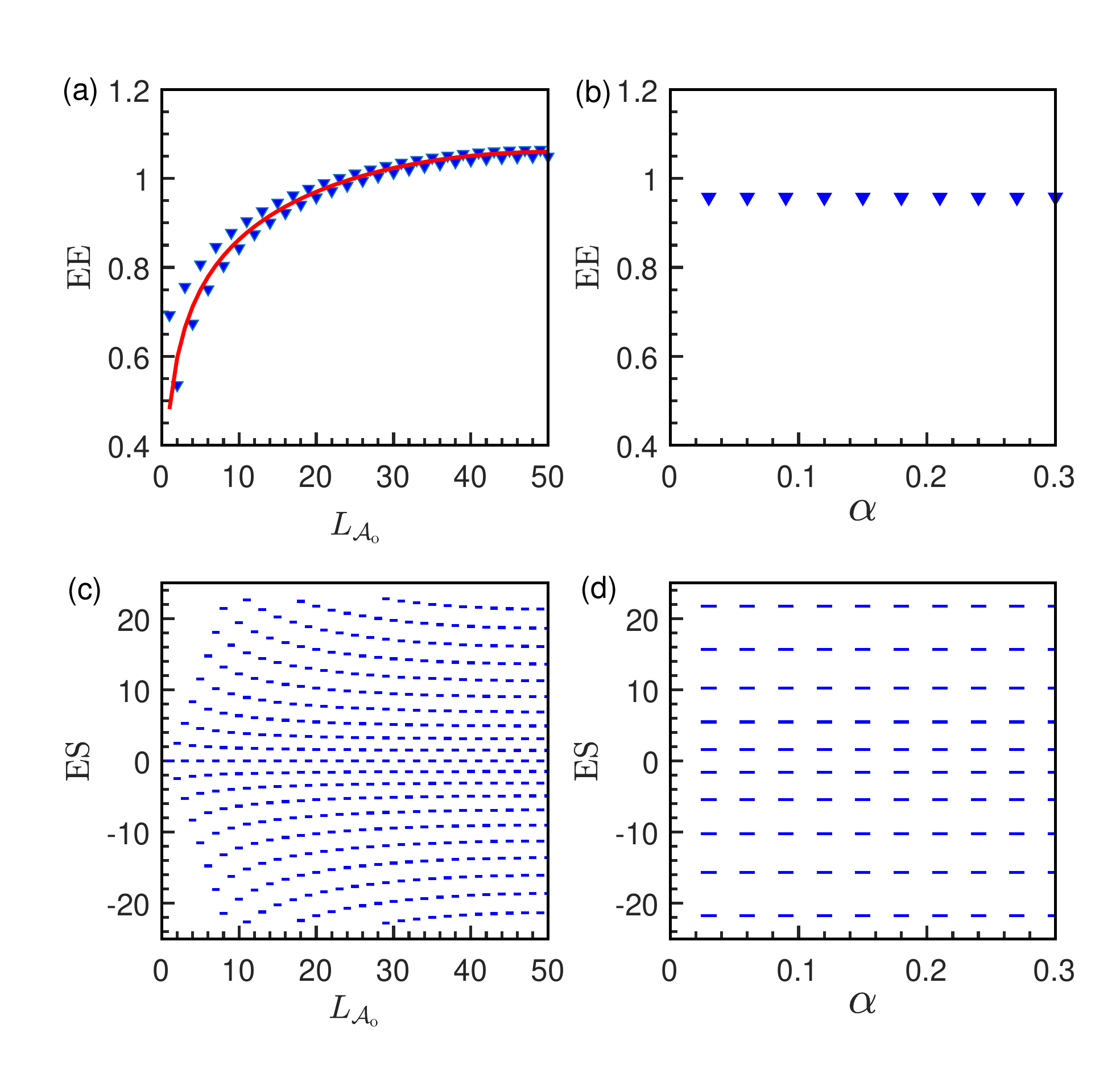}
\caption{(Color online) Entanglement entropy $S_\mathrm{EE}$ (a) \emph{v.s.} 
$L_{\mathcal{A}_\mathrm o}$ with $L=100$, $\protect\alpha=0.1$ and (b) \emph{v.s.} $%
\protect\alpha$ with $L=100$, $L_{\mathcal{A}_\mathrm o}=20$; Entanglement spectrum (c) 
\emph{v.s.} $L_{\mathcal{A}_\mathrm o}$ with $L=100$, $\protect\alpha=0.1$ and (d) \emph{%
v.s.} $\protect\alpha$ with $L=100$, $L_{\mathcal{A}_\mathrm o}=20$ for non-reciprocal
model in Eq.~(\protect\ref{nonreciprocal}). In (a), $S_{\mathrm{EE}}$
calculated via the definition in Eq.~(\protect\ref{entdef}) marked by blue
triangles is consistent with the formula of $S_\mathrm{EE}$ guided by a red
line that is inferred by the duality. (b) shows $S_\mathrm{EE}$ is
independent of $\protect\alpha$. (c) depicts EE dependence on the size of
region $\mathcal{A}_\mathrm o$. (d) shows EE is not altered when $\protect\alpha$
changes. All calculations are conducted under open boundary condition at
half-filling with $t=1$. }
\label{Fig:EEES}
\end{figure}

Remarkably, the dual Hamiltonian in Eq.~(\ref{dual1}) does not
depend on the parameter $\alpha $, which indicates that non-Hermiticity
plays no role in EE and ES in original non-Hermitian system. It is further
numerically checked in Fig.~\ref{Fig:EEES}(b) and Fig.~\ref{Fig:EEES}(d) that no changes in
EE and ES occur when we change $\alpha$. Furthermore, the quantity $\mathcal{%
OO}^{\dag }$ for the present choice is diagonal in the real-space such that
  at least one dual system is Hermitian regardless of the system partition.

In summary, the non-Hermitian non-reciprocal model in Eq.~(\ref%
{nonreciprocal}) shares the same ES and EE with the dual
Hermitian free fermi gas in Eq.~(\ref{dual1}), which exemplifies a type-I
system. This feature allows us to understand entanglement of type-I under the framework of a Hermitian system. For example, instead of complicated
calculation, we can directly extract that in the nonreciprocal model, momenta
and positions enter $S_\mathrm{EE}$ symmetrically with the expression \cite%
{1994NuPhB424443H, 2004CalabreseEntanglement, 2006CalabreseEntanglement, 2009JSPItsIFisher,
2014LeePosition} as  
\begin{equation}
S_\mathrm{EE}=\frac{1}{6}\ln \left[ \frac{L}{\pi }\sin 
\frac{\pi L_\mathcal{A}}{L}\sin \frac{\pi L_{F}}{L}\right] +\cdots~,
\end{equation}
where $L_\mathcal{A}$
 denotes the length of region $\mathcal{A}_\mathrm o$, $L_{F}$ is the
number of occupied bands of its $L$ eigenstates and $\cdots$ includes
constant and finite-size corrections from $1/L$ and higher. Figure~\ref%
{Fig:EEES}(a) shows consistence between $S_\mathrm{EE}$ from Eq.~(\ref%
{entdef}) and the duality.


\subsection{An example for Type-II: Non-Hermitian SSH model}
\label{sec:example2}

Another example is the non-Hermitian Su-Schrieffer-Heeger (SSH) model in a
bipartite lattice with $2N$ sites at half-filling, 
\begin{align}
H_{\mathrm{o}} =&\sum_{x=1}^{N-1}\left (\omega c_{2x}^{\dag }c_{2x+1}^{}+\upsilon
c_{2x+1}^{\dag }c_{2x+2}^{}\right)+\text{h.c.}  \notag \\
& +\sum_{x=1}^{N}iu\left( c_{2x}^{\dag }c_{2x}^{}-c_{2x+1}^{\dag
}c_{2x+1}^{}\right)
\end{align}%
with $u,\upsilon ,\omega \in \mathbb{R}$. Here we introduce staggered
imaginary chemical potentials. Under a periodic boundary condition (PBC),
then $H_{\text{SSH}}$ is translationally invariant and can be reformulated
in the Fock-space (i.e., momentum space) as 
\begin{equation}
H_{\mathrm{o}}\!=\!\oplus _{k}H_{k} \quad \mathrm{with}\quad H_{k}\!=\!\left[
\begin{matrix}
iu & v_{k} \\ 
v_{k}^{\ast } & -iu%
\end{matrix}
\right]~,
\end{equation}
where $v_{k}=\omega e^{-ika}+\upsilon $ with $a$ the lattice constant. The
system is $\mathcal{PT}$-invariant $\sigma _{x}H_{k}\sigma _{x}=H_{k}^{\ast }$ with
spectrum $\epsilon _{k,\pm }=\pm \sqrt{\left\vert v+we^{ika}\right\vert
^{2}-u^{2}}$ and we restrict ourselves in the region of real spectrum. 

To conduct the duality, in \textbf{Step}-\textcircled{\small\textbf{1}}, we choose the 
matrix $\mathcal{O}=\oplus _{k}\mathcal{O}_{k}$ in the momentum space to be
\begin{align}
\mathcal{O}_{k}=%
\begin{bmatrix}
-e^{-i(\theta _{k}+\varphi _{k})} & 1 \\ 
1 & e^{-i(\theta _{k}-\varphi _{k})}%
\end{bmatrix}%
~,  \label{eq_matrix_O_k}
\end{align}
where we reparametrize  
\begin{equation}
\rho _{k}e^{i\varphi _{k}}=\upsilon +\omega e^{ika}~,\quad\rho
_{k}e^{i\theta _{k}}=\epsilon _{k,+}+iu~.
\end{equation}
At half-filling, we have the two projectors 
\begin{equation}
\mathcal{P}_{\mathrm{o}}=\sum_{k\in \text{occ}}|r,k,-\rangle
\langle l,k,-|, \quad\mathcal{R}_{\mathrm{o}}=\sum_{i\in%
\mathcal{A}_\mathrm o}\sum_{s=1,2}|i,s\rangle \langle i,s|\,,
\label{PRex2}
\end{equation}%
where $\mathcal{P}_{\mathrm{o}}$ projects onto all occupied states with $%
|r,k,-\rangle $ $\left( |l,k,-\rangle \right) $ as its right(left)
eigenvector and $\mathcal{R}_{\mathrm{o}}$ defines partition with $s=1,2$
labeling two sublattices.  After a similarity transformation acting on both the two projectors in Eq.~\eqref{PRex2}, 
in \textbf{Step}-\textcircled{\small \textbf 2}, 
we interchange the roles
on positions and momenta and we have 
\begin{align}
\mathcal{P}_{\mathrm{d}}& =\sum_{k\in\mathcal{A}_\mathrm o,\ell=\pm}\mathcal{O}_{%
\mathrm{d}}^{-1}|k,\ell\rangle\! \langle k,\ell|\mathcal{O}_{\mathrm{d}}~, \\
\mathcal{R}_{\mathrm{d}}&=\sum_{x\in \mathrm{occ}}|x,-\rangle \!\langle
x,-|~,
\end{align}%
where in the dual real-space, $\mathcal{O}_{\mathrm{d}}=\oplus _{x=1}^{N}%
\mathcal{O}_{x}$ is the similarity transformation in the first step and $\ell
$ can be interpreted as internal degrees or layer indices. The expression of 
$\mathcal{O}_x$ is obtained by replacing $k$ in Eq.~(\ref{eq_matrix_O_k}) by 
$x$, and 
\begin{equation}
\rho _{x}e^{i\varphi _{x}}=\upsilon +\omega e^{i\frac{2\pi x}{Na}},\rho
_{x}e^{i\theta _{x}}=\epsilon _{x}+iu~.
\end{equation}
We can introduce the following biorthogonal eigenvectors 
\begin{equation}
|r,k,\ell\rangle=\mathcal{O}^{-1}_\mathrm{d}|k,\ell\rangle
\end{equation}
and bifermionic operators 
\begin{equation}
\psi_{r,k,\ell}^\dag|0\rangle=|r,k,\ell\rangle~,\quad
\psi_{l,k,\ell}^\dag|0\rangle =|l,k,\ell\rangle
\end{equation}
such that 
\begin{equation}
\mathcal{P}_\mathrm{d}=\sum_{k\in \mathcal{A}_\mathrm o,\ell=\pm} |r,k,\ell\rangle\langle
l,k,\ell|~,
\end{equation}
and the dual Hamiltonian takes the form as $H_\mathrm{d}=\sum_{k,s}%
\epsilon_{k,\ell}\psi_{r,k,\ell}^\dag\psi_{l,k,\ell}^{}$ where $%
\epsilon_{k,\ell}$ is the dispersion relation that is constrained by $%
\epsilon _{k,\ell }<0$ for $k\in \mathcal{A}_\mathrm o$, $\ell =\pm $. Specifically,
we can take $\mathcal{A}_\mathrm o$ to be half the chain, then the dispersion relation
can be simply chosen as $\epsilon_{k,\ell}=-2t\sqrt{N} \cos ka$. Thus, the
dual Hamiltonian can be formulated as 
\begin{equation}
H_\mathrm{d}=-t\sum_{x}\sum_{ y=x\pm a} c_x^\dag e^{i\mathbf{A}_{x,y}\cdot 
\mathbf{\sigma}+i A^0_{x,y}\sigma_0 }c_{y}^{}~,
\end{equation}
where $c_x=(c_{x,-}, c_{x,+})^\mathrm{T}$ is a two-component spinor, $%
\mathbf{\sigma}=(\sigma_x,\sigma_y,\sigma_z)$ is a vector of Pauli matrices
and $\sigma_0$ is an identity matrix. The fields $\mathbf{A}_{x,y}$ and $%
A^0_{x,y}$ reside at the link $(x,y)$ and no longer keeps anti-symmetric on
its spatial indices, $\mathbf{A}_{x,y}\not=-\mathbf{A}_{y,x},A^0_{x,y}%
\not=-A^0_{y,x}$. Thus we map non-Hermitian SSH model to non-Hermitian
non-Abelian gauge field theory. The original band indices are interpreted as
component indices and $\mathcal{R}_\mathrm{d}$ involves partition on the
internal spinor degrees as well as spatial degrees. General expressions for
the dual Hamiltonian are presented in Appendix~\ref{App:nhssh}. 
On the other hand, commutation with $\mathcal{R}_{\mathrm{o}}$ requires $%
\Theta $ to be a block diagonal matrix and no $S$ exists to make $\mathcal{O}%
^{{}}SS^{\dag }\mathcal{O}^{\dag }$ commute with $\mathcal{R}_\mathrm{o}$.
Therefore, the non-Hermitian SSH belongs to type-II, which is consistent
with its complex ES.

At the end of this section, we give some useful remarks on our duality and Dyson map. 
 Our duality is   distinct from a Dyson map \cite{2002JMP43205M,PhysRevLett89270401}. 
A Dyson map is referred to as a similarity transformation which maps
a non-Hermitian system to a Hermitian counterpart, which keeps an energy spectrum unchanged but generically   generally alters entanglement spectrum. 
In contrast, our duality has a different mission: keeping entanglement spectrum unchanged while imposing no constraints on energy spectrum. For this purpose, we have designed the duality shown in Fig.~\ref{dual2step} which includes two necessary steps. 
For a $\mathcal{PT}$-symmetric system, one can define a parity operator to act on the Hamiltonian 
just as the way as $\Theta$ in Eq.~\eqref{Theta}.
However, the factorization constraint $\Theta=\mathcal{O}\mathcal{O}^\dagger$ does not allow $\Theta$ to be such a parity operator.
In fact, our duality is applicable to a general non-Hermitian system since as shown in Fig.~\ref{dual2step} the duality targets at invariance of ES and EE and no restriction is needed for energy spectra.

\section{Conclusions}
\label{section_conclusions} 
In this paper, we are interested in the role
that non-Hermiticity plays in quantum entanglement and develop a rigorous
duality for probing the role. We make an initial step towards a unified
picture of non-Hermiticity and Hermiticity from the entanglement
perspective. Explicitly, we have considered non-Hermitian non-interacting systems whose   Hamiltonians are
assumed to act on the Hilbert space with a complete set of biorthonormal eigenvectors  
and  possess an entirely real spectrum. We classify these systems into
two types, 	which is summarized in Table~\ref{Tab-1}. 
For type-I, non-Hermiticity plays no role and thus we can
efficiently obtain entanglement entropy and entanglement spectrum by means of well-studied results in Hermitian
systems. For type-II, non-Hermiticity indeed plays an intrinsic role.
Several implications and applications are discussed.

Motivated by this work,
we present here several questions for future study. First, is there a
similar/generalized duality or generalized LU for characterizing many-body
entanglement of non-Hermitian interacting systems \cite%
{2019arXiv191101590XClassification, 2020arXiv200610278Y}? For example, it is important  to define non-Hermitian version of LRE states via generalized LU. Second,
is it possible to find Wannier interpolation on non-Hermitian non-interacting
entanglement \cite{2015PhRvBLeeFree}? Such an interpolation may help us to
semi-analytically understand ES of non-Hermitian fermion systems. Third, how
can we further physically distinguish two subclasses of type-II systems? In
Corollary 1, we have shown that ES of type-II systems may be either complex
or real. So, it is interesting to further investigate finer structures of
type-II systems. Third, a free Hamiltonian plays a role as a mean-field
theory of an interacting system. What is the relation between two
interacting systems if their mean-field theory are dual to each other?
	Fourth, it is appealing to generalize our duality into momentum space entanglement \cite{2014PhRvL113y6404L, 2016PhRvL117a0603D}.
Furthermore, it is important to perform experimental measurement to
distinguish entanglement behaviors of the two-type non-Hermitian systems
using the experimental breakthroughs \cite{2001PhRvA..64e2312J,
2015Natur.528...77I} and in particular to confirm our statement for type-I
non-Hermitian systems.
After the first arxiv version, we were aware that there are some other interesting investigations on entanglement of non-Hermitian systems, such as Refs.~\cite{arxiv_entangle1,arxiv_entangle2,arxiv_entangle3}. In the future, it will be interesting to combine these increasing new findings and duality together.
\section*{Acknowledgments}

This work was supported in part by the Sun Yat-sen University startup grant,
Guangdong Basic and Applied Basic Research Foundation under Grant No. 2020B1515120100, National Natural Science Foundation of China (NSFC)
Grant (No. 11847608 \& No. 12074438).

\newpage
\begin{appendix}
In the appendix, we make some explanation on the choices of $\mathcal O $ in the first step of the duality
and detailed derivations on the dual model for non-Hermitian SSH model.

\section{A two-site model}
\label{App:strans} 

In this appendix, We concentrate  on the procedures to determine the type of a given non-Hermitian system. 
As is shown in Fig.~\ref{dual2step}, our duality is split into  two steps. 
In Step-\textcircled{\small \bf 1}, a similarity transformation $\mathcal O$ are conducted 
on both Fock-space and real-space projectors $\mathcal P_\mathrm o$ and $\mathcal R_\mathrm o$ 
and then in Step-\textcircled{\small \bf 2}  an interchange between interpretation on momenta and positions follows.
According the duality, we classify non-Hermitian systems into two type. Explicitly, 
if at least one Hermitian dual system $H_\mathrm d$ exists, a non-Hermitian system belongs to type-I. Otherwise, it belongs to type-II.
In practice, we need to check all possible $\mathcal O$ in Step-\textcircled{\small \bf 1}. 

Here we consider a system with only two lattice sites that reads 
\begin{equation}
H_{\mathrm{o}}=re^{i\theta }c_{1}^{\dag }c_{1}^{{}}+sc_{1}^{\dag
}c_{2}+re^{-i\theta }c_{2}^{\dag }c_{2}^{{}}+sc_{2}^{\dag }c_{1}.
\end{equation}%
where $c_{s}\left( s=1,2\right) $ is a fermion annihilation operator at the
site $s$. The system has only one particle. 
Non-Hermiticity arises from a complex-valued chemical potential.
Following the preliminary, we introduce bifermionic operators 
\begin{equation}
\begin{split}
\psi _{l-}^{{}} &=\frac{1}{\sqrt 2\cos\alpha}(e^{i\alpha} c_{1}+ c_{2})~, \\
\psi _{l+}^{{}} &=-\frac{1}{\sqrt 2\cos\alpha}( - c_{1}+e^{i\alpha}c_{2})~, \\
\psi _{r-}^{\dag } &= \frac{1}{\sqrt 2}(c_{1}^{\dag }+ e^{-i\alpha} c_{2}^{\dag })~, \\
\psi _{r+}^{\dag } &=\frac{1}{\sqrt 2} (-e^{-i\alpha }c_{1}^{\dag }+c_{2}^{\dag })~,
\end{split}%
\end{equation}
with $se^{i\alpha }=ir\sin \theta +\sqrt{s^{2}-r^{2}\sin ^{2}\theta }$. 
And then we have the spectral decomposition 
\begin{equation}
H_{\mathrm{o}}=\epsilon_{-}\psi _{r-}^{\dag }\psi _{l-}^{{}}+\epsilon_{+}\psi _{r+}^{\dag }\psi _{l+}^{{}}  ~,
\label{AppHo1}
\end{equation}%
where $\epsilon_{\pm }=r\cos \theta \pm \sqrt{s^{2}-r^{2}\sin ^{2}\theta }$
are eigenenergies of the two states\ $|r,\pm \rangle =\psi _{r\pm }^{\dag
}|0\rangle $. The ground state describes occupation of the state $|r,-\rangle 
$ and the corresponding Fock-space projector $\mathcal{P}_{\mathrm{o}}$ is 
\begin{equation}
\mathcal{P}_{\mathrm{o}}=\psi _{r-}^{\dag }|0\rangle \langle
0|\psi _{l-}^{{}}~.
\end{equation}%
We make a partition where the subregion $\mathcal{A}_\mathrm o$ only contains the
first site with the real-space projector being
\begin{equation}
\mathcal{R}_{\mathrm{o}}=c_{1}^{\dag }|0\rangle \langle
0|c_{1}^{{}}~.
\end{equation}%
Let's start our duality. In \textbf{Step}-\textcircled{\small \bf 1}, we can choose $\mathcal{O}$ to
be 
\begin{equation}
\mathcal{O}=\frac{1}{\sqrt 2}e^{i\alpha }c_{1}^{\dag }|0\rangle \langle
0|c_{1}^{{}}+\frac{1}{\sqrt 2} c_{1}^{\dag }|0\rangle \langle 0|c_{2}^{{}}-\frac{1}{\sqrt 2}c_{2}^{\dag
}|0\rangle \langle 0|c_{1}^{{}}+\frac{1}{\sqrt 2}e^{i\alpha }c_{2}^{\dag }|0\rangle \langle
0|c_{2}^{{}}~,
\end{equation}%
which defines a similarity transformation on both $\mathcal{P}_{\mathrm{o}}$
and $\mathcal{R}_{\mathrm{o}}$. We introduce the notations%
\begin{equation}
\begin{split}
c_{\pm } &=\mathcal{O}^{-1}\psi _{r\pm }\mathcal{O}, \\
c_{\pm }^{\dag } &=%
\mathcal{O}^{-1}\psi _{l\pm }^{\dag }\mathcal{O}~, \\
\psi _{r1,2} &=\mathcal{O}^{-1}c_{1,2}\mathcal{O},\\
 \psi _{l1,2}^{\dag } & =%
\mathcal{O}^{-1}c_{1,2}^{\dag }\mathcal{O}~,
\end{split}
\label{dualoperator}
\end{equation}
under which we get a compacted form after action of $\mathcal{O}$  
\begin{equation}
\mathcal{O}^{-1}\mathcal{P}_{\mathrm{o}}\mathcal{O}=c_{-}^{\dag }|0\rangle
\langle 0|c_{-},\quad \mathcal{O}^{-1}\mathcal{R}_{\mathrm{o}}\mathcal{O}=\psi
_{r1}^{\dag }|0\rangle \langle 0|\psi _{l1}^{{}}~.
\end{equation}%
It is easy to check that in Eq.~\eqref{dualoperator}, $c_{\pm }$ and $c_{\pm }^{\dag }$ are conventional
fermion operators while $\psi _{r1,2}$ and $\psi _{l1,2}^{\dag }$ are
bifermionic operators [See Sec.~\ref{section_duality_preli}]. In \textbf{Step}-\textcircled{\small \bf 2}, we exchange  indices $%
\left\{ \pm \right\} $ and spatial indices $\left\{ 1,2\right\} $, thus
obtaining two projectors $\mathcal{P}_{\mathrm{d}}$ and $\mathcal{R}_{%
\mathrm{d}}$ in the dual system%
\begin{equation}
\mathcal{P}_{\mathrm{d}}=\psi _{r-}^{\dag }|0\rangle \langle 0|\psi _{l-}^{{}},\quad
\mathcal{R}_{\mathrm{d}}=c_{1}^{\dag }|0\rangle\langle 0|c_{1}~,
\label{App:dualRP}
\end{equation}%
where 
\begin{equation}
\begin{split}
\psi _{l-}^{{}} &=\frac{1}{\sqrt 2\cos \alpha }(  c_{1}-e^{-i\alpha } c_{2})~, \\
\psi _{l+}^{{}} &=\frac{1}{\sqrt 2\cos \alpha } ( e^{-i\alpha }c_{1}+c_{2})~,\\
\psi _{r-}^{\dag } & =\frac{1}{\sqrt 2}(e^{i\alpha}c_{1}^{\dag }-c_{2}^{\dag }) ~,\\
\psi _{r+}^{\dag }&=\frac{1}{\sqrt 2}(c_{1}^{\dag }+e^{i\alpha}c_{2}^{\dag })~.
\end{split}
\end{equation}
One should not confuse  operators in Eq.~(\ref{App:dualRP}) with those
in the original system $H_{\mathrm{o}}$. We can build the dual
Hamiltonian $H_{\mathrm{d}}$\ by designing its spectrum $\pm \cos \alpha $, 
\begin{align}
H_{\mathrm{d}} &=-\cos\alpha \psi _{r-}^{\dag }\psi _{l-}^{{}}+\cos \alpha
\psi _{r+}^{\dag }\psi _{l+}^{{}} \notag \\
&=-i\sin\alpha  c_{1}^{\dag }c_{1}^{{}}+c_{1}^{\dag }c_{2}+
c_{2}^{\dag }c_{1}+i \sin \alpha c_{2}^{\dag }c_{2}^{{}}
\end{align}%
When $\alpha \not=0$, non-Hermiticity of $H_{\mathrm{d}}$ arises from the
complex-valued chemical potential. Consistently,  the quantity $\Theta=\mathcal{OO}^{\dag }$ in Eq.~\eqref{Theta}
\begin{equation}
\Theta=2c_{1}^{\dag }|0\rangle \langle 0|c_{1}+2i\sin
\alpha c_{1}^{\dag }|0\rangle \langle 0|c_{2}+2i\sin \alpha c_{2}^{\dag
}|0\rangle \langle 0|c_{1}+2c_{2}^{\dag }|0\rangle \langle 0|c_{2}
\end{equation}%
indeed fails to commute with $\mathcal{R}_{\mathrm{o}}$%
\begin{equation}
\left[ \Theta ,\mathcal{R}_{\mathrm{o}}\right] =-2i\sin \alpha c_{1}^{\dag
}|0\rangle \langle 0|c_{2}+2i\sin \alpha c_{2}^{\dag }|0\rangle \langle
0|c_{1}
\end{equation}%
To determine the type of $H_{\mathrm{o}}$, we have to check whether at least
one Hermitian dual $H_{\mathrm{d}}$ exists. Suppose 
\begin{equation}
S=\lambda
_{1}c_{1}^{\dag }|0\rangle \langle 0|c_{1}+\lambda _{2}c_{2}^{\dag
}|0\rangle \langle 0|c_{2}~.
\end{equation}
 Then $\mathcal{O}S$ also satisfies the
requirement in {\bf Step}-\textcircled{\small \bf 1}. However, we can not find any $S$ to make $\left[ 
\mathcal{O}SS^{\dag }\mathcal{O}^{\dag },\mathcal{R}_{\mathrm{o}}\right] =0$, 
\begin{equation}
\left[ \mathcal{O}SS^{\dag }\mathcal{O}^{\dag },\mathcal{R}_{\mathrm{o}}%
\right] =\left( e^{-i\alpha }\left\vert \lambda _{1}\right\vert
^{2}-e^{i\alpha }\left\vert \lambda _{2}\right\vert ^{2}\right) c_{1}^{\dag
}|0\rangle \langle 0|c_{2}-\left( e^{-i\alpha }\left\vert \lambda
_{1}\right\vert ^{2}-e^{i\alpha }\left\vert \lambda _{2}\right\vert
^{2}\right) c_{2}^{\dag }|0\rangle \langle 0|c_{1}~.
\end{equation}%
Therefore, $H_{\mathrm{o}}$ belongs to type-II. On the other hand, we can
directly calculate the reduced density matrix 
\begin{equation}
\rho _{\mathrm o}=\mathrm{Tr}_{\mathcal{B}_\mathrm o}|r,-\rangle \langle l,-|=\frac{1}{2\cos
\alpha }\left( e^{-i\alpha }c_{1}^{\dag }|0\rangle \langle
0|c_{1}+e^{i\alpha }|0\rangle \langle 0|\right) 
\end{equation}%
and entanglement spectrum is complex-valued and 
\begin{equation}
S_{\mathrm {EE}}=-\frac{
e^{-i\alpha }}{2\cos \alpha }\log \frac{e^{-i\alpha }}{2\cos\alpha}-\frac{e^{i\alpha}}{2\cos\alpha}\log \frac{e^{i\alpha }}{%
2\cos \alpha }~.
\end{equation}

As comparison, we consider non-reciprocal model on a lattices with two sites,%
\begin{equation}
H_{\mathrm{o}}=rc_{1}^{\dag }c_{1}+t_{12}c_{1}^{\dag }c_{2}+rc_{2}^{\dag
}c_{2}+t_{21}c_{2}^{\dag }c_{1}  \label{AppHo2}
\end{equation}%
with single particle and subregion containing the first site. In step-1, we can
choose $\mathcal{O}$ to be 
\begin{equation}
\mathcal{O}=\sqrt{t_{12}}c_{1}^{\dag }|0\rangle \langle 0|c_{1}^{{}}+\sqrt{%
t_{12}}c_{1}^{\dag }|0\rangle \langle 0|c_{2}^{{}}-\sqrt{t_{21}}c_{2}^{\dag
}|0\rangle \langle 0|c_{1}^{{}}+\sqrt{t_{21}}c_{2}^{\dag }|0\rangle \langle
0|c_{2}^{{}}~.
\end{equation}%
To determine the type of Hamiltonian in Eq.~(\ref{AppHo2}), suppose $%
S=\lambda _{1}c_{1}^{\dag }|0\rangle \langle 0|c_{1}+\lambda _{2}c_{2}^{\dag
}|0\rangle \langle 0|c_{2}$ and then 
\begin{equation}
\left[ \mathcal{O}SS^{\dag }\mathcal{O}^{\dag },\mathcal{R}_{\mathrm{o}}%
\right] =\left( \left\vert \lambda _{1}\right\vert ^{2}-\left\vert \lambda
_{2}\right\vert ^{2}\right) \sqrt{t_{12}t_{21}}c_{1}^{\dag }|0\rangle
\langle 0|c_{2}-\left( \left\vert \lambda _{1}\right\vert ^{2}-\left\vert
\lambda _{2}\right\vert ^{2}\right) \sqrt{t_{12}t_{21}}c_{2}^{\dag
}|0\rangle \langle 0|c_{1}~.
\end{equation}%
When $\left\vert \lambda _{1}\right\vert =\left\vert \lambda _{2}\right\vert 
$, $\left[ \mathcal{O}SS^{\dag }\mathcal{O}^{\dag },\mathcal{R}_{\mathrm{o}}%
\right] =0$ and we are allowed to build a map to a Hermitian system. When $%
\left\vert \lambda _{1}\right\vert \not=\left\vert \lambda _{2}\right\vert $%
, the dual system is non-Hermitian. Therefore, $H_{\mathrm{o}}$ in Eq. (\ref
{AppHo2}) belongs to type-I. One can also follow the steps in Fig.~\ref{dual2step} 
to construct the dual system, which
is just like what we do for the system in Eq. (\ref{AppHo1}).

\section{Non-Hermitian SSH model}

\label{App:nhssh} 
Here we present details on duality of non-Hermitian SSH
model discussed in Sec.~\ref{section_two_typical_examples}. To be more transparent, we work in the framework of Dirac's
notations. The similarity transformation $\mathcal{O}=\oplus_k \mathcal{O}%
_{k}$ can be written as 
\begin{align}
\mathcal{O}_{k}=&-e^{-i\left( \theta _{k}+\varphi _{k}\right) }|k,1\rangle
\langle k,1|+|k,1\rangle \langle k,2|  +|k,2\rangle \langle k,1| +e^{-i\left( \theta _{k}-\varphi _{k}\right)
}|k,2\rangle \langle k,2|~,
\end{align}%
where $|k,s\rangle $ denotes a basis vector carrying momentum $k$ on
sublattice $s$. So we have $\mathfrak{B}$, 
\begin{align}
\mathfrak{B} &=\sum_{x\in  \mathcal{A}}\mathcal O^{-1}|x,s\rangle \langle x,s|\mathcal O  \notag \\
&=\frac{1}{N}\sum_{x\in \mathcal{A}}\sum_{k,k^{\prime }}\mathcal O_{k}^{-1}|x,s\rangle \langle
x,s|\mathcal O_{k'}  \notag \\
&=\frac{1}{N}\sum_{x\in  \mathcal{A}}\sum_{k,k^{\prime }}\frac{e^{i\left( k-k^{\prime
}\right) x}}{1+e^{-2i\theta _{k}}}[\left( e^{i\varphi _{k}-i\varphi
_{k^{\prime }}}e^{-i\theta _{k}-i\theta _{k^{\prime }}}+1\right) |k,1\rangle
\langle k^{\prime },1|  \notag \\
&+\left( -e^{i\varphi _{k}}e^{-i\theta _{k}}+e^{i\varphi _{k^{\prime
}}}e^{-i\theta _{k^{\prime }}}\right) |k,1\rangle \langle k^{\prime },2|
\notag +\left( e^{-i\varphi _{k}}e^{-i\theta _{k}}-e^{-i\varphi _{k^{\prime
}}}e^{-i\theta _{k^{\prime }}}\right) |k,2\rangle \langle k^{\prime },1|
\notag \\
&+\left( e^{-i\varphi _{k}+i\varphi _{k^{\prime }}}e^{-i\theta _{k}-i\theta
_{k^{\prime }}}+1\right) |k,2\rangle \langle k',2|]~,
\end{align}%
where we use the relation 
\begin{align}
\mathcal{O}_{k}|x,1\rangle &=\frac{1}{\sqrt{N}}e^{ikx}\left( -e^{-i\left(
\theta _{k}+\varphi _{k}\right) }|k,1\rangle +|k,2\rangle \right) \\
\mathcal{O}_{k}|x,2\rangle &=\frac{1}{\sqrt{N}}e^{ikx}\left( |k,1\rangle
+e^{-i\left( \theta _{k}-\varphi _{k}\right) }|k,2\rangle \right)
\end{align}%
with $\rho _{k}e^{i\varphi _{k}}=\upsilon +\omega e^{ika}$ and $\rho
_{k}e^{i\theta _{k}}=\epsilon _{k}+iu$. Exchange the roles of positions and
momenta and we have the dual Fock-space projector $\mathcal{P}_{\mathrm{d}}$
to occupied states 

\begin{align}
\mathcal{P}_{\mathrm{d}} &=\frac{1}{N}\sum_{x,y,k\in  \mathcal{A}}\frac{e^{i\left(
x-y\right) k}}{1+e^{-2i\theta _{x}}}[\left( e^{i\varphi _{x}-i\varphi
_{y}}e^{-i\theta _{x}-i\theta _{y}}+1\right) |x,-\rangle \langle y,-|  \notag
\\
&+\left[ -e^{i\varphi _{x}}e^{-i\theta _{x}}+e^{i\varphi _{y}}e^{-i\theta
_{y}}\right] |x,-\rangle \langle y,+|+\left[ e^{-i\varphi _{x}}e^{-i\theta
_{x}}-e^{-i\varphi _{y}}e^{-i\theta _{y}}\right] |x,+\rangle \langle y,-|
\notag \\
&+\left[ e^{-i\varphi _{x}+\varphi _{y}}e^{-i\theta _{x}-i\theta _{y}}+1%
\right] |x,+\rangle \langle y,+|~.
\end{align}%
Therefore, we have the dual Hamiltonian, 
\begin{equation}
H_\mathrm{d}=\frac{1}{N}\sum_{x,y,k,\ell,\ell^{\prime }}\epsilon _{k,\ell
}e^{i\left( x-y\right) k}U_{xy}^{\ell \ell ^{\prime }}c_{x,\ell }^{\dag
}c_{y,\ell ^{\prime }}^{}~,
\end{equation}%
where%
\begin{align}
U_{xy}=&\frac{1}{1+e^{-2i\theta _{x}}}    
\begin{bmatrix}
e^{i\varphi _{x}-i\varphi _{y}}e^{-i\theta _{x}-i\theta _{y}}+1 & 
-e^{i\varphi _{x}}e^{-i\theta _{x}}+e^{i\varphi _{y}}e^{-i\theta _{y}} \\ 
e^{-i\varphi _{x}}e^{-i\theta _{x}}-e^{-i\varphi _{y}}e^{-i\theta _{y}} & 
e^{-i\varphi _{x}+i\varphi _{y}}e^{-i\theta _{x}-i\theta _{y}}+1%
\end{bmatrix}%
\end{align}
and $\rho _{x}e^{i\varphi _{x}}=\upsilon +\omega e^{i\frac{2\pi x}{Na}},\rho
_{x}e^{i\theta _{x}}=\epsilon _{x}+iu $. The spectrum dispersion relation is
constrained to $\epsilon _{k,\ell }<0$ when $k\in \mathcal{A}$. Its form depends on
choices of region $\mathcal{A}_\mathrm o$. For example, if $\mathcal{A}_\mathrm o$ is half the chain, we can take $%
\epsilon _{k,\ell }=-2t\sqrt{N}\cos ka$ to be independent of index $\ell$.
In this case, we arrange $c_x=(c_{x,-},c_{x,+})^T$ as a two-component spinor
such that  the dual Hamiltonian can be formulated compactly,
\begin{align}
H_{\mathrm{d}}&=- t \sum_x \sum_{y=x\pm a}c_x^\dagger U_{xy}c_y^{}  =-t \sum_x \sum_{y=x\pm a} c_x^\dagger e^{i \mathbf{A}_{x,y}\cdot \sigma+i
A_{x,y}^0\sigma_0}c_y^{}~.
\end{align}
Here we identify $U_{xy}=e^{i\mathbf{A}_{x,y}\cdot \mathrm{\sigma}%
+iA_{x,y}^0\sigma_0}$ to give non-Abelian gauge field theory of the
non-Hermitian version. The form of $\epsilon_{k,\ell}$ for a general region $%
\mathcal A_\mathrm o$ can be obtained by adding a proper chemical potential.
\end{appendix}

\bibliographystyle{SciPost_bibstyle}

\nolinenumbers
\end{document}